\documentclass[referee]{raa}
\usepackage{graphicx,times}

\begin{document}

\title{Time resolved spectral analysis of the prompt emission of long gamma ray bursts with GeV Emission}

   \volnopage{Vol.0 (200x) No.0, 000--000}      
   \setcounter{page}{1}   

\author{A.R. Rao 
  \inst{1}
\and Rupal Basak
  \inst{1}
\and J. Bhattacharya
  \inst{2}
\and S. Chandra
  \inst{2}
\and N. Maheshwari 
  \inst{3}
\and M. Choudhury
  \inst{4}
\and Ranjeev Misra
  \inst{5}
 }

\institute{
Tata Institute of Fundamental Research, Mumbai, India \\
\and 
Indian Institute of Technology, Kanpur, India \\
\and
Indian Institute of Technology, Mumbai, India \\
\and
Centre for Excellence in Basic Sciences, Mumbai, India \\
\and
Inter-University Centre for Astronomy and Astrophysics, Pune, India \\
}

\date{Received~~2009 month day; accepted~~2009~~month day}

\abstract{
We make a detailed time resolved spectroscopy of bright long gamma ray bursts
(GRBs) which show significant GeV emissions   (GRB~080916C, GRB~090902B, and 
GRB~090926A). 
In addition to the standard Band model, we also use a model consisting
of a blackbody and a power-law to fit the spectra. We find that for
the latter model there are indications for an additional soft component
in the spectra.  While previous studies have shown that such models are
required for GRB~090902B, here  we find that a composite spectral model
consisting of
two black bodies and a power law adequately fit the data of all the three bright
GRBs.
We investigate the
evolution of the spectral parameters and find several  generic  interesting
features for
all three GRBs,
like a) temperatures of the black bodies are strongly correlated to each other,
b) flux in the black body components are strongly correlated to each other,
c) the temperatures of the black body trace the profile of the individual 
pulses of the GRBs, and d) the characteristics of the power law component
like the spectral index and the delayed onset bear a close similarity to the
emission characteristics in the GeV regions.  
We discuss the implications of these results to  the 
possibility of identifying the radiation mechanisms during the prompt emission
of GRBs.
}

\authorrunning{A. R. Rao et al. }            
\titlerunning{Time resolved spectral analysis of long gamma ray bursts }

\maketitle

\section{Introduction}

Gamma-ray bursts (GRBs) are some of the highest energetic events detected,
with the emissions spanning over several decades of energy, from a few tens of
keV to tens of GeVs (see, for eg. Abdo et al. 2009a). The origin of the
bursting mechanism as well as the radiative processes that give rise to the 
emission are still a matter of intense debate (Zhang 2007; Dado \& Dar 2009).
Although there is a great variety in the shapes of the light curves observed,
the bursts are historically classified into short and long as per the duration
of the prompt emission of which the former has duration of less than 2 seconds
while the long GRBs may last from a few seconds  to hundreds of seconds, but
recently there have been efforts to take the diverse spectral, temporal
properties as well as the location into account and classify them into Type I
and Type II classes (Zhang 2007; Zhang et al. 2009).
Both types show afterglows in lower energy bands (spanning from X-rays to radio)
that may last from days to weeks (see Gehrels et al. 2009, for a general review of GRBs).

The approach to investigate the working mechanism of the GRB involves two
independent paradigms: first, the theoretical assumption and/or the simulation
of a central engine along with the processes that may lead to the outburst
and/or the subsequent emission processes followed by phenomenological fitting of
the data by the assumed models (Goodman 1986; Paczynski 1986; Dar 2006; Dado et al.
2007; King 2007; Falcke \& Biermann 1995; Metzger et al. 2011; Zhang \& Meszaros 2002), and second, the data driven analysis of processes
(Band et al. 1993; Amati et al. 2002; Ackermann et al. 2010). 
Presupposition of a theoretical scenario may at times induce a bias in the
analysis process and the subsequent interpretation of the data, while the data
driven reasoning may at times lead to either empirical or unphysical
explanations.

Long GRBs provide the opportunity to analyse the time-resolved spectra of the
prompt emission with a comparative statistical advantage over the short GRBs. The
challenge of such an analysis lies in fitting the data with a physically
meaningful model, in contrast to the Band model (Band et al. 1993) which
provides a very good statistical fit to the data with two power law components
smoothly joining at a peak energy. 
In an attempt to fit with more  physically meaningful models
Ryde (2004; 2005) and Ryde \& Pe'er  (2009) have fitted the time-resolved
spectra of clear, fast rise exponential decay (FRED) pulses of BATSE GRBs with a
blackbody and a power law. The data quality of the detectors were not suitable
for more nuanced analysis. Recently many attempts have been made to mimic the
Band model by more physically relevant models (Ackermann et al. 2011), whilst
there also have been efforts to extend the Band model into the very high energy
gamma-ray as well as the less then 50 keV X-ray regime of the electromagnetic
spectrum (Abdo et al. 2009b).

The observational analysis of GRBs received a boost with the launch of 
Fermi satellite observatory. The Fermi Gamma-ray Space Telescope hosts two
instruments, the Large Area Telescope (LAT, 20 MeV to more than 300
GeV, Atwood et al. 2009) and the Gamma-ray Burst Monitor (GBM, 8 keV -- 40
MeV, Meegan et al. 2009), which together are capable of measuring the spectral
parameters of GRBs across seven decades in energy.
One of the remarkable observations from the Fermi satellite is the detection
of high energy ($\sim$GeV) emission from GRBs. Detection of photons up to the
energy of 30 GeV constrained the Lorentz factor of the jet to be greater than
$\sim$1000  (Abdo et al. 2009a; Ghirlanda et al. 2010a);
 detection of GeV photons from the short GRB~090510 helped put stringent 
limits on the violation of Lorentz invariance (Abdo et al. 2009c). Though the
detection of a large number of photons above 100 MeV ($>$100) in some of the
bright GRBs like GRB~090902B helped in making a detailed time resolved
spectroscopy, no unified spectral model explaining the prompt emission of GRBs
across the full energy range has yet emerged. For example, GRB~090902B shows a
separate peaked Band emission in the 50 keV -- 1 MeV region and a power-law connects
the $>$ 100 MeV emission (Abdo et al. 2009b) whereas GRB~080916C shows a single
Band function fitting across the full energy range (Abdo et al. 2009a). 
A detailed time-resolved study of 17 GRBs with high energy emission showed the
possibility of five spectral combination to explain the data
(Zhang et al. 2011).

Ryde et al. (2010) have found that the time resolved spectra of GRB~090902B
does not agree with the physically meaningful model of a black body and a
power-law, but a continuous distribution of temperature fits the data. 
 The need for such more complex models to fit
the data may lead the way to a better understanding of the radiative processes
responsible for the
prompt emission and to distinguish different dynamical models. However, it may
also be possible
that such complexities exist only for certain GRBs and hence inferences drawn from
such studies may not
be general. Indeed,  Ryde et al. (2010) suggest 
that in other GRBs like GRB~080916C, the thermal component may be lacking. It is
important to know whether a
single empirical model can fit the time-resolved spectra of all GRBs that have
sufficiently good signal to noise data. 

In this paper we report the results of time resolved spectral analyses of
three long GRBs which had shown intense GeV emissions and which  are also bright
 in the 1 keV -- 10 MeV range  (fluence $>$10$^{-4}$ erg cm$^{-2}$).
Our aim is to find a spectral model which can be used to describe the
time-resolved spectral data and connect the spectral parameters to the high
energy emission.
Our approach is data driven, our interpretation is phenomenological and  our
attempt is to look for an empirical model that fits the time resolved spectra
of all bright GRBs. 
In section 2, we describe the data we used for our analysis, and the softwares
we used for our purposes. In section 3, we  describe the details of the
time resolved spectral analysis of the light curves of the three GRBs. We
present our results in section 4, followed by a discussion on the results.

\section{Data Selection And Extraction}

Zhang et al. (2011) have made a systematic time resolved spectral analysis
of  a complete sample of 17 GRBs with Fermi LAT detection (another 6 are 
added to this list based on a systematic search of Fermi-LAT data base
using a matched filter technique - see Zheng et al. 2012). The complete details of
all these GRBs are also given in Ackermann et al.(2013).  We have selected
three long GRBs from this list which are bright in MeV 
(fluence $>$ 10$^{-4}$ erg cm$^{-2}$) and GeV regions ($>$100 photons 
above 100 MeV). Only one other GRB in this list has intense GeV emission,
but it is a short GRB with comparatively lower fluence (GRB~090510 with a
fluence of 2 $\times$ 10$^{-5}$ erg cm$^{-2}$ and duration of 0.3 s).
These three GRBs show a delayed onset of $>$ 100 MeV emission (see
also Ackermann et al. 2011; Abdo et al. 2009a,b,c).

Fermi satellite has two detectors namely, Gamma ray Burst Monitor (GBM) and
Large Area Telescope (LAT). GBM is the primary instrument for the detection and
study of GRB prompt emission. It detects X-rays and low energy $\gamma$-rays. It has two scintillation detectors: the sodium iodide (NaI) detector is
sensitive in the $\sim$ 8 keV to $\sim$ 900 keV range while the BGO energy
range is $\sim$ 200 keV to $\sim$ 40 MeV (Meegan et al. 2009). The other
primary detector on board Fermi is LAT. It has a large field of view, such that
it can see 20\% of the entire sky at any  time, and over a period of 3 hours it
scans the whole sky. The effective area of LAT is 9500 cm$^2$. For both the
detectors, we used the data which were publicly available at the Fermi mission
website\footnote{http://fermi.gsfc.nasa.gov/ssc/data/access}.

\begin{table}
\caption{GRB coordinates,   trigger time, and the GBM detector numbers.}
\centering
\begin{tabular}{c c c c c c c}
\hline\hline
GRB & RA (J2000) & Dec (J2000) & Trigger Time & NaI & BGO \\
\hline
080916C & 07h59m23s & -56$^{\circ}$38'20.1" & 00:12:45.61 UT & 3,4 & 0 \\
090902B & 17h38m00s & +27$^{\circ}$19'26.6" & 11:05:08.31 UT  & 0,9 & 1\\
090926A & 23h33m36s & -66$^{\circ}$19'25.9" & 04:20:41.00 UT & 3,7 & 1 \\
\hline
\end{tabular}
\label{table:GRBlocations}
\end{table}

We used the standard procedure for the GBM analysis, closely following the
method described in Basak \& Rao (2012a,b). 
For each GRB we used two or more  NaI detectors with high detected count rates,
according to
the data present in the "Time Tagged Event"  (TTE) file.
We chose one BGO detector depending on the selected NaI detectors. 
If the NaI detectors 1-6 were selected we used BGO 0, else we used BGO
1. For ambiguous cases, we used the BGO which showed a higher count rate. 
We used the TTE file to extract the light curves and spectra. 
Using the spectral analysis tool RMFIT (version 3.3pr7), developed by user
contributions  of Fermi Science Support Center (FSSC), we created the time bins
from the original TTE file to reduce the fluctuations. 
After binning, we fitted a linear or a cubic polynomial to the background, by
choosing time intervals before and after the prompt emission phase. 
For the time  resolved spectral analysis, we followed the method of Basak \& Rao (2013)
and
selected time bins with fixed excess counts: we repeated the analysis with a total excess counts
of $\sim$2000 and $\sim$4000 for GRB~090902B. Since we got consistent results with these count rates,
we used an excess count of $\sim$2000 for the analysis of the other two GRBs. 
We grouped the spectral files, the response files and the background files such
that in each spectral bin sufficient number of counts  are available for good
statistics (typically 40 counts for NaI detectors and 50 -- 60 counts for the
BGO detectors) and analyzed the data with  the spectral analysis software XSPEC
(Version 12.7.0).

For analyzing the LAT data, we used the LAT ScienceTools--v9r23p1 package. 
We also used the "transient" response function. We have considered the time
periods for which the signal to noise ratio is considerably high, whereas the
data lost due to inadequate signal is not significant. 
We chose the LAT data based on the time and position measured by other detectors
like GBM.
We downloaded the data from the weekly Fermi-LAT database from NASA HEASARC
website, using the precise GRB  coordinates available in the literature. 
To filter out the emission coming from the earth's atmosphere due to cosmic
rays, we used a maximum zenith angle cutoff value of 105$^\circ$. We binned the
data in time using the tool gtbin provided by the NASA ScienceTools package. We
chose the
energy range from 100 MeV - 300 GeV, and obtained the light curves. For the
spectral
analysis, we found that the uncertainties in the data beyond 2 GeV were large.
We considered the energy range from 30 MeV--2 GeV for spectral fitting. 

The coordinates, trigger time, and the GBM detectors used for the analysis are
listed in Table 1, for the three GRBs studied here.

\section{Time resolved spectral analysis}
GRB~090902B is the brightest of the three selected GRBs and we first 
attempt a time-resolved spectral analysis of this GRB.
The light curve of the prompt emission has multiple peaks in all the energy
bands (see for example Abdo et al. 2009b; Ackerman et al. 2013) with two prominent
peaks. The LAT counts are delayed by about  3 seconds
 compared to the counts in the GBM. The time resolved spectra has been fit by
Abdo et al. (2009b) where they have extended the Band model with a power law
that extends to below 50 keV and above to very high energy gamma-rays (in the
GeV regime). 
Zhang et al. (2011) found that the Band model becomes increasingly narrower
with smaller time intervals and they conclude that a black body and a power-law
gives a correct description of the time resolved spectra, though other models 
like cutoff power law with a power-law too could fit the data. They, however,
concluded that the black body and a power-law is unique to this source and there
is no evidence for such a combination in other GRBs. Ryde et al. (2010)
could get acceptable fits to the time resolved spectra using a composite model
involving a
multi-color black body and a power law. The blackbody temperature is a
continuous distribution with the flux at each temperature having a power-law 
relation with temperature up to a maximum temperature, T$_{max}$.

\begin{figure}\centering
{
\includegraphics[scale = 0.7, angle=0]{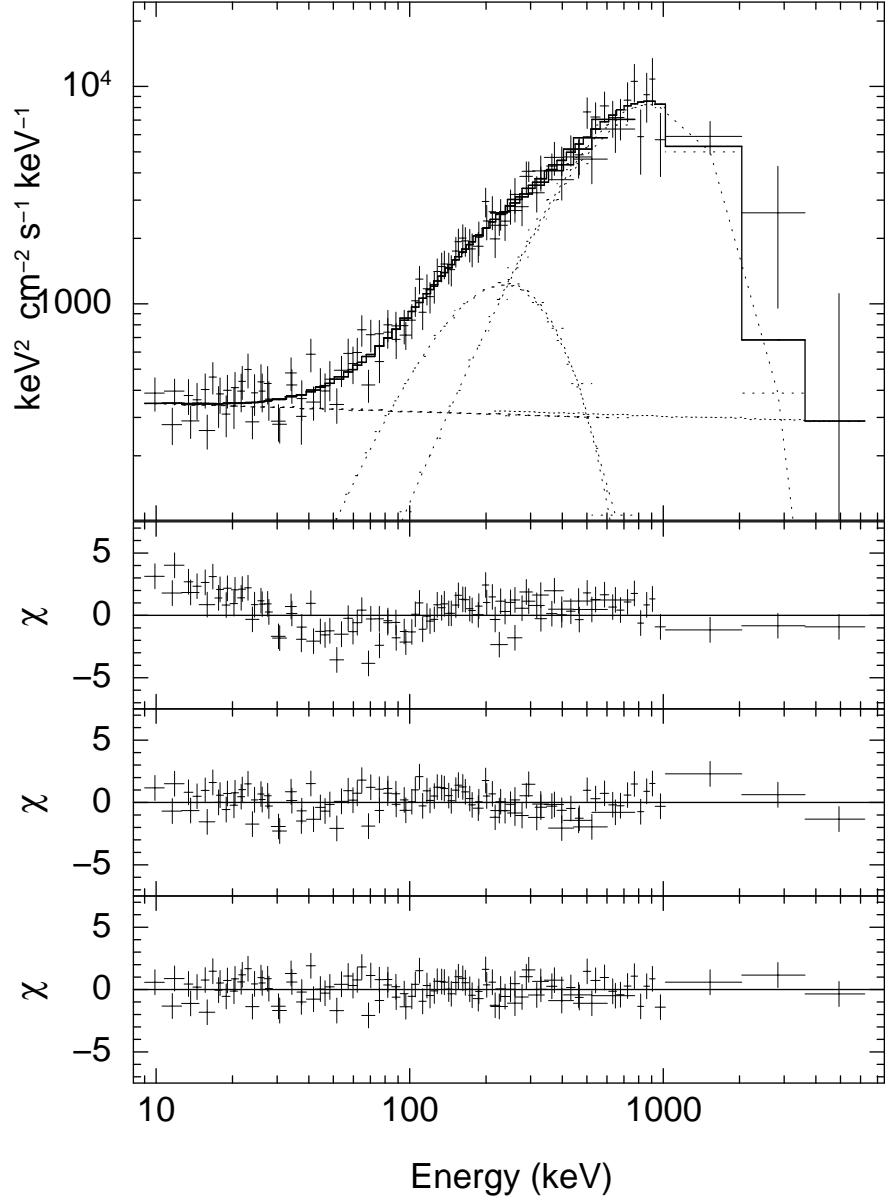}
}
\caption{Unfolded energy spectrum of a time resolved spectrum of GRB~090902B
using
a model consisting of two black bodies and a power-law (2BBPL model) 
shown in the top panel. Individual model components are shown as
dotted lines. The residuals to the fit are shown in the bottom three panels 
for (successively from the top) the Band model, black body and a power-law (BBPL
model) and the  2BBPL model.}
\label{fig1}
\end{figure}

\begin{figure}\centering
\includegraphics[scale = 0.7, angle=0]{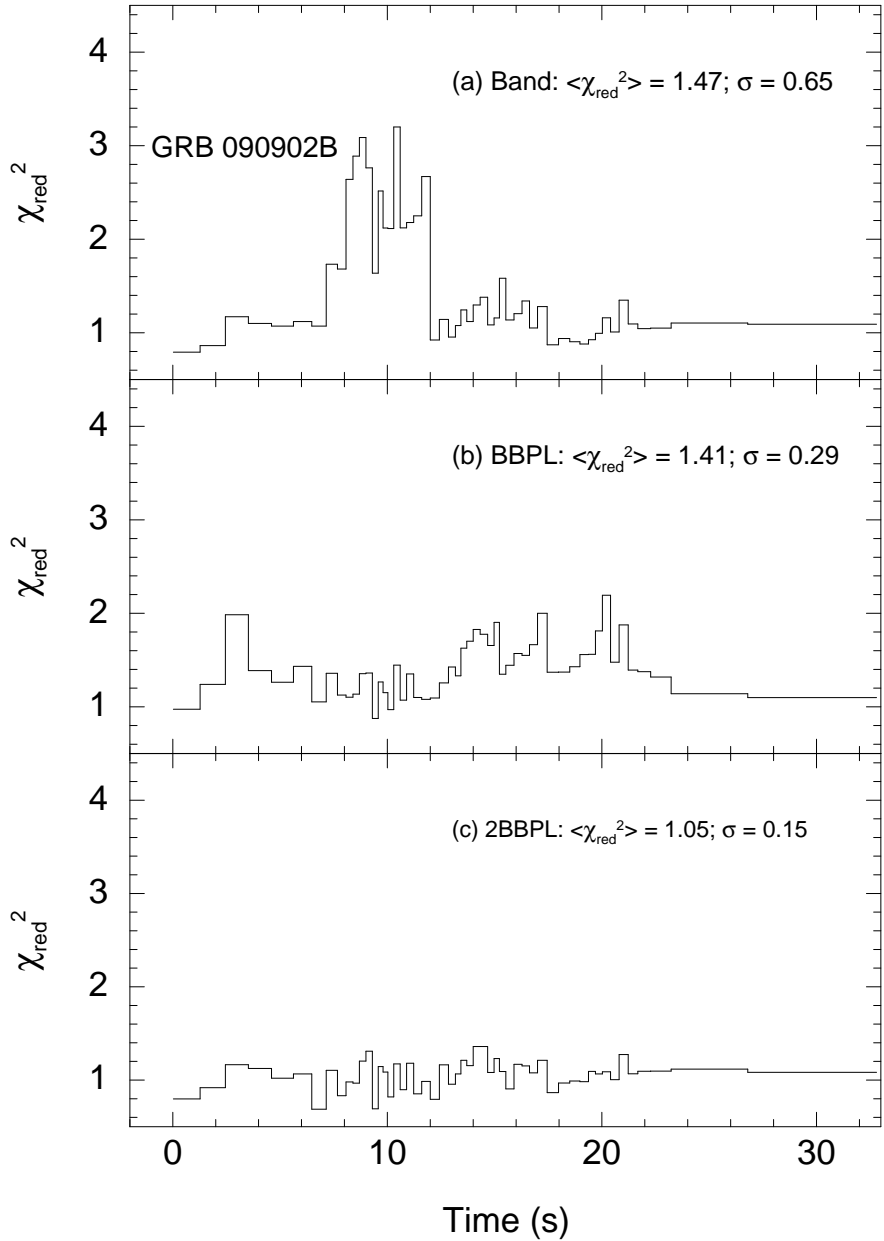}
\caption{The values of reduced $\chi^2$ for a time-resolved spectral analysis
of GRB~090902B for (a) Band model, (b)   black body and  power-law model (BBPL)
and (c) two black bodies and a power-law (2BBPL).  The average values and the rms
deviation
in them ($\sigma$) are also indicated in each of the panels.}
\label{fig2}
\end{figure}

In our attempt to arrive at an acceptable time resolved spectral model for 
the bright GRBs, we first fit the time resolved GBM data (with $\sim$2000 counts
per spectrum)
with the ``Blackbody + Powerlaw'' model (hereafter referred to as the
BBPL model). We found high values for the reduced $\chi^2$ ($\chi_r^2$),
particularly
during the rising part of the pulse features. 
In  Figure \ref{fig1} we show the representative example of one of the spectra.
When we fit only with a Band model, we get a high $\chi^2$, which improves for 
the BBPL model, but there still are residuals, particularly at the
peak of the spectrum. When we include another blackbody 
(the lower blackbody with a temperature  kT $\sim$ a few tens of keV, while
the higher blackbody with kT $>$ 100 keV) along with a power law (hereafter 
referred to as the  2BBPL model), we find a very significant improvement in
the value of $\chi_r^2$. 
The improvement to the fit is shown graphically in Figure \ref{fig2} for the
time resolved spectral analysis. The average value of $\chi_r^2$ is 1.47 for 
the Band model fitting (with a dispersion of 0.65) and it improves to 1.41 for 
a black body with a power-law. The addition of another black body
component improves
the $\chi_r^2$ to 1.05 (average value) with a dispersion of 0.15. The
 power-law index $\Gamma$
has an average value of 1.76 with a dispersion of 0.17 (shown in the bottom panel of 
Figure 3).
When we use a continuous distribution of temperature (hereafter called the
mBBPL model) to the time resolved
data (as was done by Ryde et al. 2010), we find that data
are consistent with this model ($\chi_r^2$ = 1.14 with a dispersion of 0.15). 
We find that the LAT spectra are fit satisfactorily by a   power law
model with the power law index obtained from the GBM time resolved analysis.
Zhang et al. (2011) report a LAT power law index of 1.76 for this GRB.

\begin{figure}
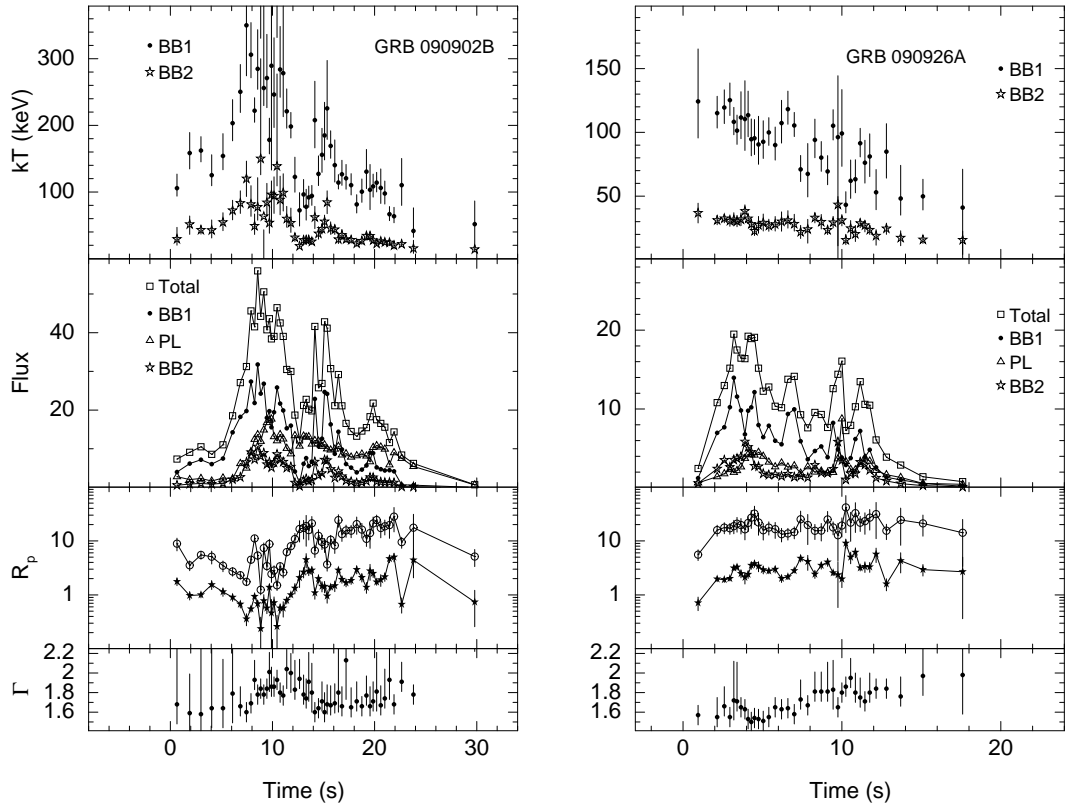
\centering
{
\begin{tabular}{lc}
\includegraphics[scale=0.45, angle=0]{090902_kT_flux_rev1.ps}
\includegraphics[scale=0.45, angle=0]{090926_kT_flux_rev1.ps}
\end{tabular}
}
\caption{The variation of the parameters for a time resolved spectral analysis of GRB~090902B 
using the 2BBPL model is shown in the left panels of the figure. The panels,
successively from top, show the temperatures of the two black body components; 
fluxes (in units of 10$^{-6}$ erg cm$^{-2}$ s$^{-1}$)  in various
components; the dimensionless photospheric radius parameter R$_p$ (see text) in units
of 10$^{-19}$ for the higher black body (stars) and lower blackbody (open circles); and
the power law index. Similar quantities for GRB~090926A are shown in the right 
panels of the figure.}
\label{fig3}
\end{figure}

\begin{figure}\centering
{
\begin{tabular}{lc}
\includegraphics[scale=0.31, angle=-90]{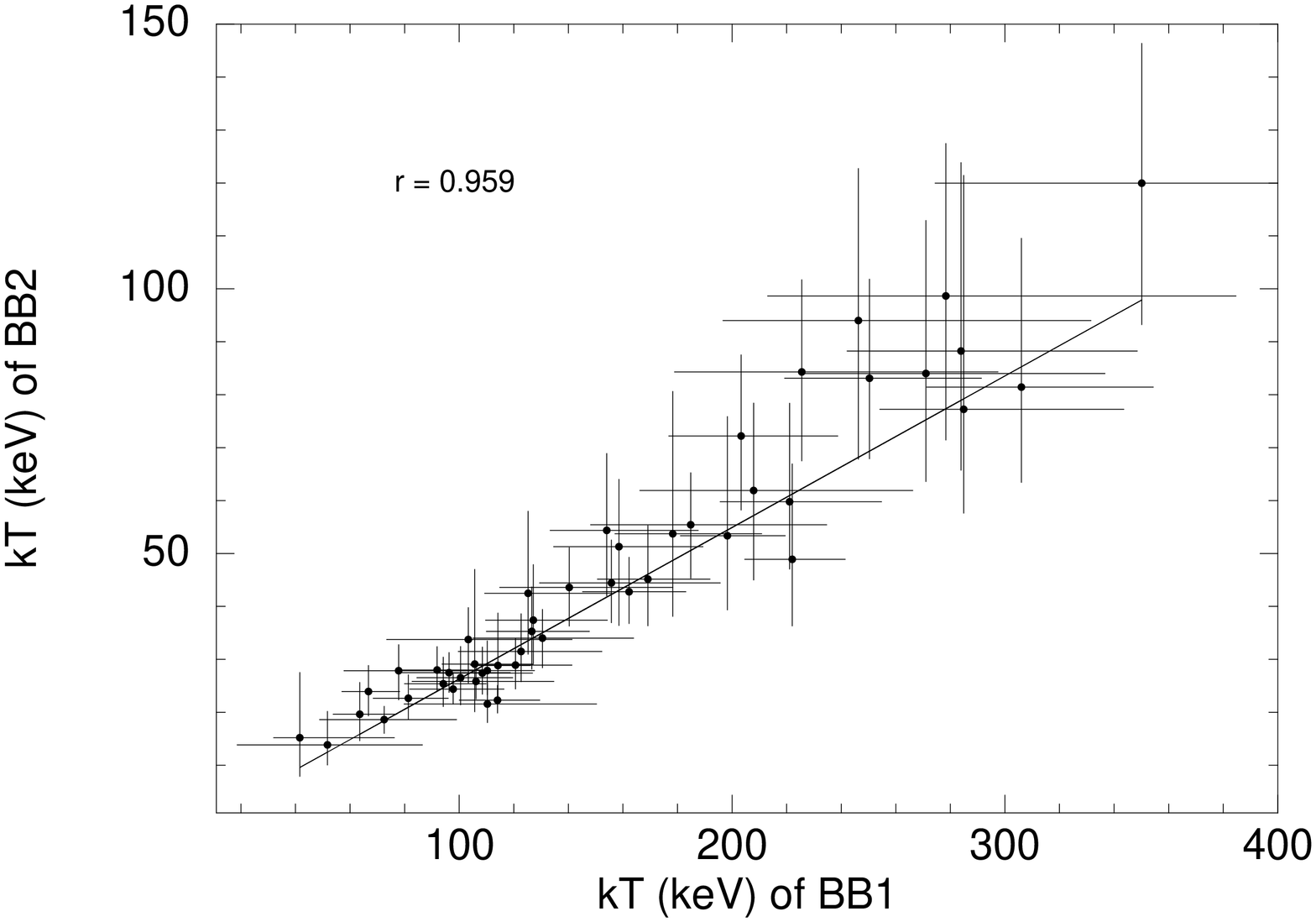}
\includegraphics[scale=0.31, angle=-90]{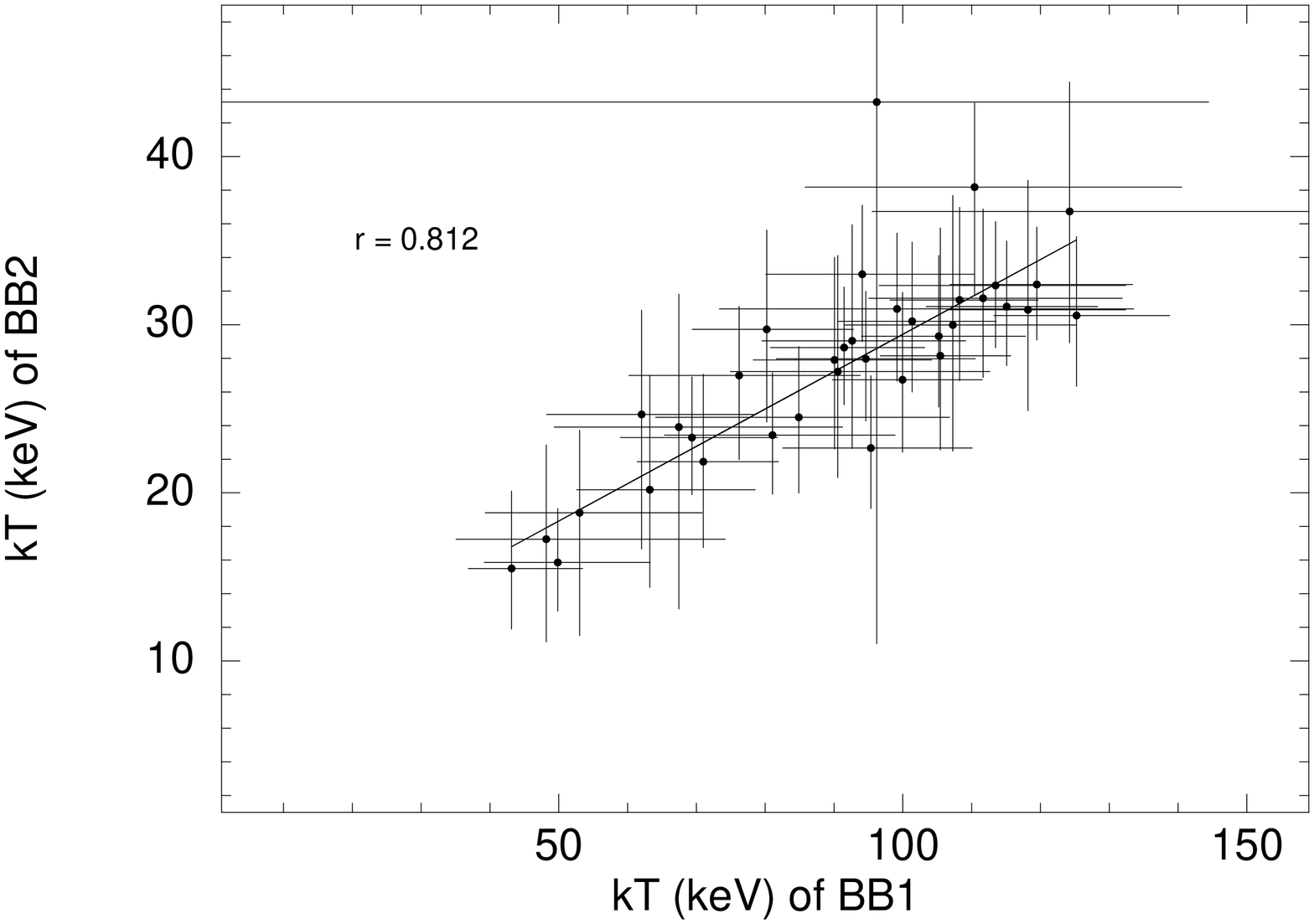}
\end{tabular}
}
\caption{The scatter diagram of the temperature (kT) of the higher blackbody
component
with the temperature (kT) of the lower black body for GRB~090902B (left panel)
and GRB~090926A
 (right panel) respectively. The correlation coefficients, r, are indicated
in the panels. The straight lines show linear fits to the data.}
\label{fig5}
\end{figure}

The evolution of spectral parameters for GRB~090902B is shown in Figure
\ref{fig3}, left panels.
The temperatures of the two black bodies are plotted in the top panel and 
the flux values in the individual components are shown in the next panel.
The most interesting observation here is the evolution of the two temperatures,
which follows exactly the same trend, suggesting that a single phenomenon is
driving their evolution.
In Figure \ref{fig5}, left panel, a scatter diagram of the individual
temperatures of the two black bodies are shown.
A good level of correlation (correlation coefficient, r, of 0.96 for 48 data
points) is seen  and a slope of 0.29$\pm$0.04  is obtained.
The normalisation of the two black bodies, too, are correlated to each other.

Following Ryde \& Pe'er (2009) we have defined a dimensional photospheric
radius parameter R$_p$ as 
\begin{equation}
 R_p = (\frac{F_{BB}}{\sigma T^4})^{1/2}
\end{equation}
where F$_{BB}$ is the black body flux, T is the temperature of the black body and
$\sigma$ is the Stefan–Boltzmann constant. R$_p$ is proportional to
the photospheric emission radius for a GRB of given redshift and 
Lorentz factor (see equations 3 and 4 of Ryde \& Pe'er 2009). 
R$_p$ is plotted in Figure 3, for both the black bodies of the 2BBPL model.
Since the temperatures and fluxes are correlated with a similar fraction (about 0.3),
R$_p$ for the low temperature black body is a factor of 6 higher than that of 
the high temperature black body. Though the GRBs considered here do not have
clear pulse structures, an increasing trend of R$_p$ can be seen in the figure.

The light curve of the prompt emission of GRB~090926A has multiple peaks in all
the energy bands (see for example Ackermann et al. 2011). 
The LAT counts are delayed by about 5 seconds compared to the counts in the GBM.
The time resolved spectra have been fit by Ackermann et al. (2011) where they
have extended the Band model with a cutoff-power law, reporting a spectral break
at around 1.4 GeV, while claiming that this additional component is more
prominent than the Band component.
In our attempt to fit the time resolved spectra  (taken with total source counts
of $\sim$2000 in each time bin)
with the BBPL model, again we were unable to find a proper fit, whereas
two black bodies (the lower blackbody with a temperature  kT $\sim$ few tens of
keV, while the higher blackbody with a kT $>$ 100 keV) along with a power law did
provide adequate and acceptable fit statistics  for this source too.

In Figure Fig 3 (right panels) the temperature and flux evolutions for GRB~090926A are shown.
The LAT spectra are fit satisfactorily by a   power law model with the power law
index obtained from the GBM data, though Zhang et al. (2011) report a power-law
index of 2.03 for this source in the LAT energy range.
Again, the change in the power law flux  appears to drive the LAT light curve.
For this source too the most interesting observation  is the evolution of the
two temperatures, which  follow exactly the same trend, suggesting that a single
phenomenon is driving their evolution. A scatter plot of the two temperatures
are shown in Figure \ref{fig5}, right panel.
A good level of correlation (correlation coefficient, r, of 0.81 for 36 data
points) is seen  and a slope of 0.22$\pm$0.06  is obtained. The normalization of
the blackbody components are also correlated to each other.


\begin{table}
\caption{The reduced $\chi^2$ values for the various models}
\centering
\begin{tabular}{l l l l l}
\hline\hline
GRB & Band  & BBPL  & mBBPL & 2BBPL \\
\hline
080916C &  1.05 & 1.14 & 1.07 & 1.04 \\
090902B (0 -- 7.2 s)  &  1.19 & 1.38 & 1.10 & 1.11 \\
090902B (7.2 -- 12 s)  &  3.81 & 1.25 & 1.15 & 1.16 \\
090902B (12  -- 35.2 s)  &  1.33 & 1.65 & 1.24 & 1.22 \\
090926A & 1.11 & 1.68 & 1.19 & 1.15 \\
\hline
\end{tabular}
\label{table:GRBlocations}
\end{table}

For GRB~080916C the LAT light curve
reached its peak a few seconds after the trigger, whereas the GBM light curve
reached the peak immediately after the trigger (Abdo et al. 2009a).
The delay between the GBM and LAT counts for this source is about 4 seconds.
For the time resolved spectral study we  find that
the Band model provides a better fit 
compared to the BBPL model.
In our attempt to identify an uniform spectral distribution for diverse GRBs, we
followed the method of Basak \& Rao (2013) and
did an uniform analysis for all the three GRBs. We selected three episodes in 
GRB~090902B (0 -- 7.2 s, 7.2 -- 12 s, and 12 -- 35.2 s, respectively -- see Figure 2) and the 
other two GRBs. We made a simultaneous fit to the time resolved spectral files with
the following constraints: a) indices $\alpha$ and $\beta$ tied for the Band model
b) power-law index tied in the BBPL model, c)  power-law index and the temperature variation index
tied in the mBBPL model and d) power-law index and the ratio of temperatures
and normalizations tied in the  2BBPL model. The resultant reduced $\chi^2$ values are shown in
Table 2. We can conclude that 2BBPL model gives an uniformly good fit to the data for
all the three GRBs.



\subsection{Evolution of the power-law flux}

An  examination of the flux evolution for all the three GRBs shows that 
the power-law flux has a delayed start and the black body temperature and 
flux decrease sharply towards the end of the burst. To investigate the evolution
of the flux beyond the prompt emission, we have obtained the spectral data 
with long integration times in 4 bins for GRB~090902B (25 -- 30 s, 30 -- 40 s,
40 -- 60 s, and 60 -- 100 s, respectively), 3 bins for GRB~090926A (17 -- 30 s,
30 -- 50 s, and 50 -- 70 s, respectively) and 1 bin for GRB~080916C (64 -- 100
s).
We fit a power-law to the GBM data with the value of index frozen at the 
average values obtained from the time resolved analysis of the prompt emission.
We investigate below whether the power-law flux in the GBM range (reflecting
the non-thermal part of the prompt emission) relates to the LAT flux
(which is assumed to be of non-thermal origin).

\begin{figure}\centering
\includegraphics[scale=0.70]{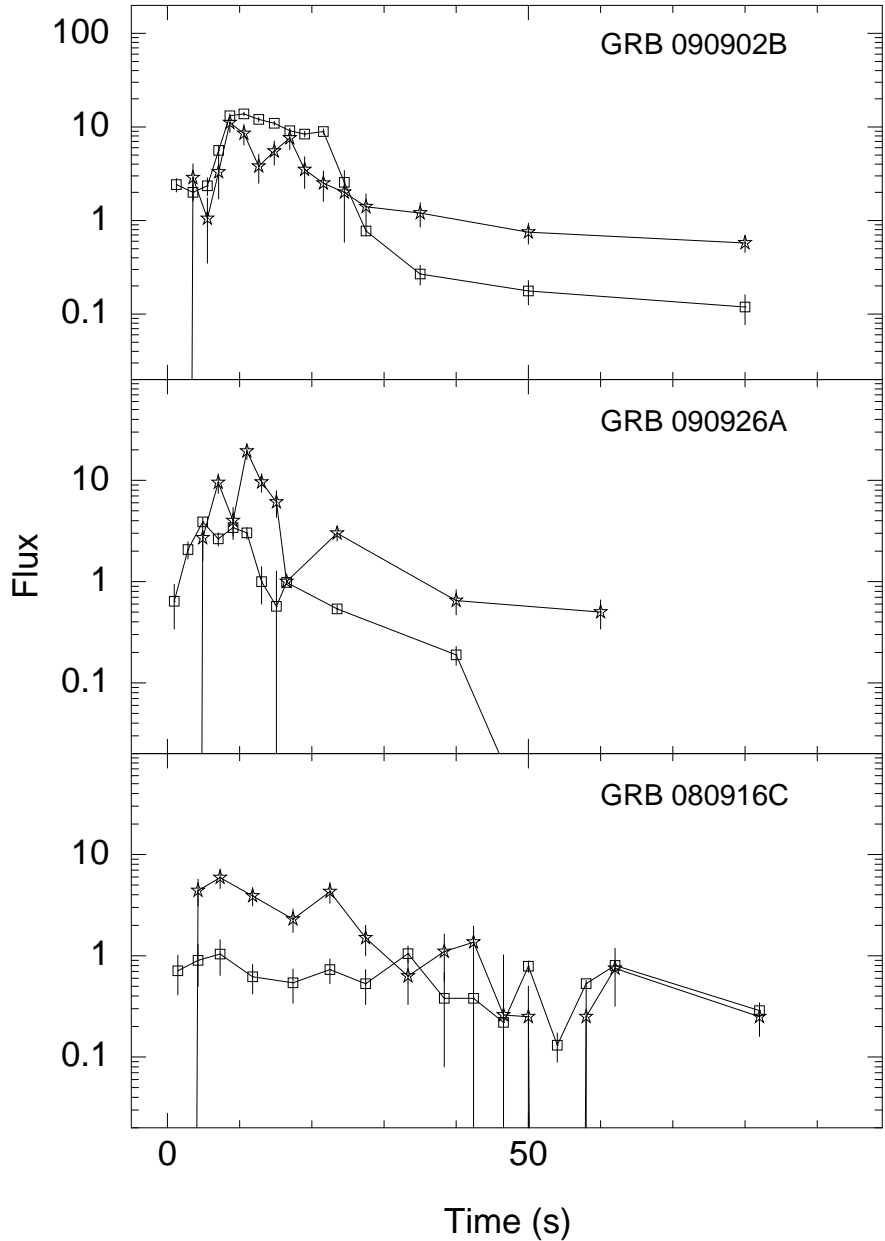}
\caption{ The variation of the power-law flux (in the units
 of 10$^{-6}$ erg cm$^{-2}$ s$^{-1}$) for the time resolved spectral 
analysis using the  2BBPL model (open squares) and
the LAT flux (shown as stars, in units of the observed LAT count rates for
events above 100 MeV)
shown as a function of time for GRB~090902B (top panel), GRB~090926A (middle
panel), and GRB~080916C (bottom panel).
}
\label{fig6}
\end{figure}

\begin{figure}\centering
\includegraphics[scale=0.7, angle=0]{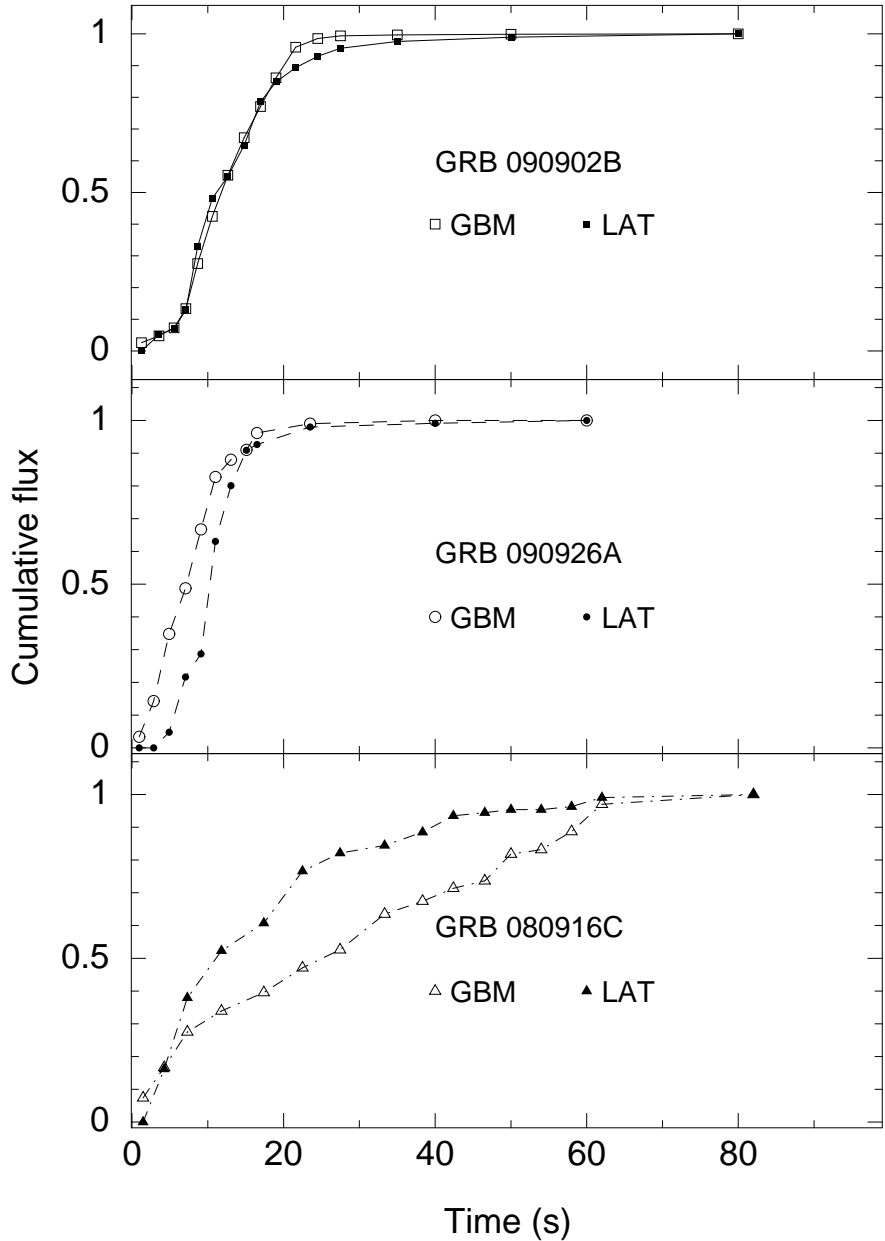}
\caption{The cumulative integrated flux distribution for GBM power-law flux
(open symbols) shown
along with a similar distribution for the LAT flux (filled symbols) for 
GRB~090902B (top panel), GRB~090926A (middle
panel), and GRB~080916C (bottom panel). 
}
\label{fig7}
\end{figure}

In Figure \ref{fig6} we show the evolution of the power-law flux (shown as open
boxes, in the units of 10$^{-6}$ erg cm$^{-2}$ s$^{-1}$)  along with the LAT
flux (shown as stars, in the units of LAT count rates  $>$100 MeV).
It is quite fortuitous that the choice of these units make the quantities have a
similar range of values.
The LAT bore-sight angle for these three GRBs are quite  similar to each other
(49$^\circ$, 50$^\circ$, and 47$^\circ$, respectively for GRB~080916C, 
GRB~090902B, and GRB~090926A -- see Zheng et al. 2012) and hence the observed count
rates can be deemed as the relative LAT fluxes for a given GRB.
It can be seen that the power-law flux tracks the LAT flux quite well. 
For GRB~090902B, the power-law flux is a factor of 10 lower than the peak flux
in the initial 6 seconds after the trigger and the rise thereafter coincides
quite smoothly with the rise in the LAT flux.
The first peak (at 9 -- 11 s) in the two energy ranges too coincide with each
other, though the fall after 20 s is much steeper for the power-law flux.
In GRB~090926A, on the other hand, the power-law flux, though delayed ($\sim$3
s) rises earlier than the LAT flux ($\sim$5 s).
In GRB~080916C the two fluxes track each other remarkably well, including a dip
in both the fluxes at $\sim$55 s after the trigger.
Cumulative flux distributions which highlight the similarities in the trends
during the rising phase of each GRB are shown in Figure \ref{fig7}.

\begin{figure}\centering
\includegraphics[scale=0.7]{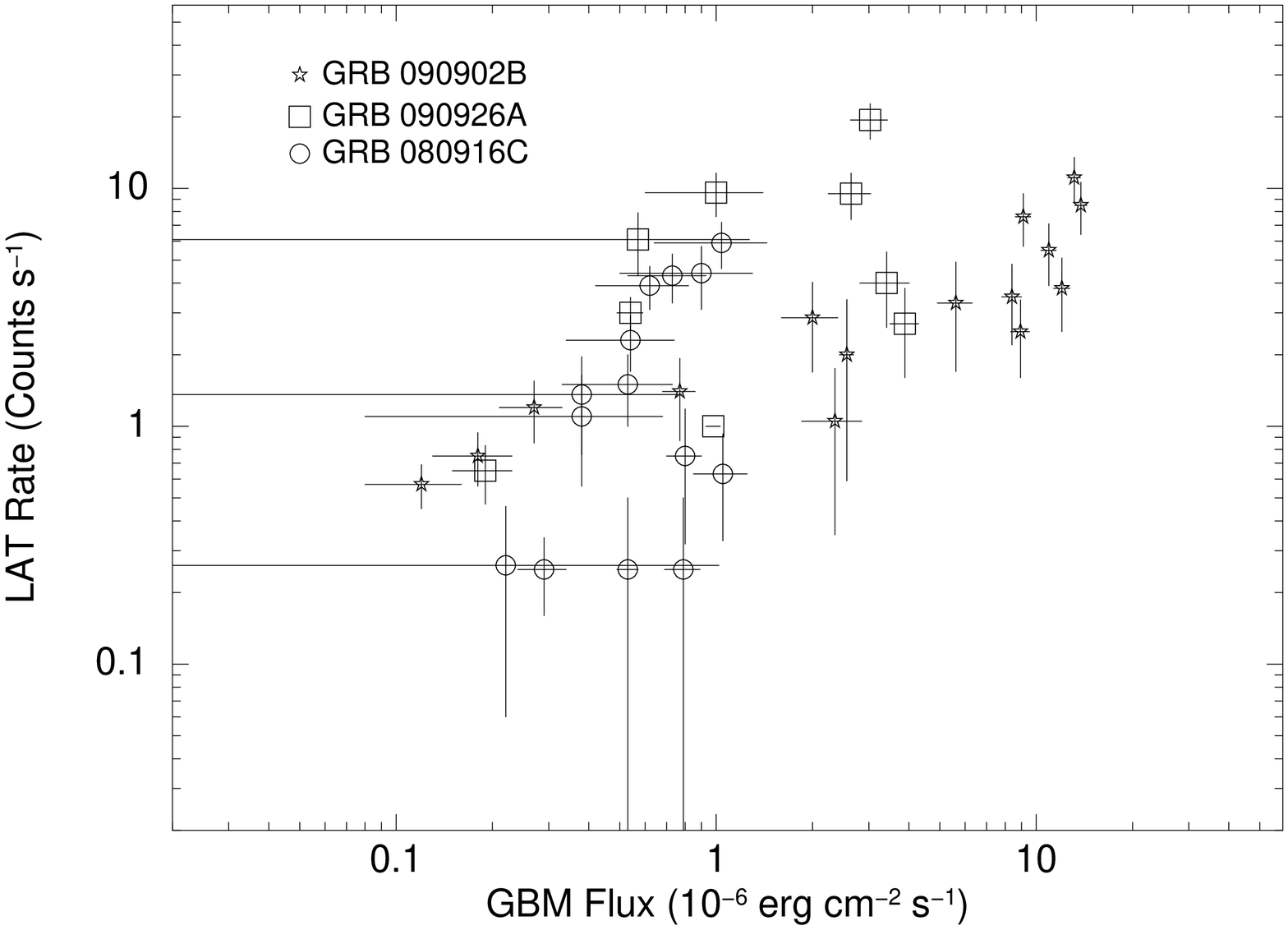}
\caption{A scatter diagram of the power-law flux in the  2BBPL model with the
LAT count rates for a time resolved analysis of GRB~090902B (stars), GRB~090926A
(squares),
and GRB~080916C (circles).}
\label{fig8}
\end{figure}

In Figure \ref{fig8} we show a scatter plot of the LAT flux against
GBM power-law flux for all the three GRBs. The two fluxes are correlated
very well for GRB~090902B (correlation coefficient, r,  of
0.84
for 16 data points)  
and weakly correlated
for the other two GRBs (r of 0.32 and 0.36, respectively for GRB~090926A and GRB~080916C).

\section{Discussion and Conclusions}

The time integrated as well as the time resolved  spectra  of GRBs are
traditionally
fit with the Band model, in which two power-laws smoothly join together
(see eg. Nava et al. 2011; Ghirlanda et al. 2010b).
The existence of power-laws immediately points towards some non-thermal
phenomena, but the variations in the Band spectral parameters with time, for a
given GRB, is quite difficult to reconcile with any reasonable physical scenario
of non-thermal radiation mechanisms (see for example Ghirlanda et al. 2003).
There have been attempts to model the time-resolved spectra of GRBs with some
sort of photospheric thermal emissions (Ryde 2004; 2005), but
the variations of the parameters like  the black body temperature with time
shows some increase in the initial parts which is quite difficult to reconcile
with any physical scenario.

In the present work we have demonstrated that a model consisting of two black
bodies and a power-law adequately fit the time resolved data. We have further shown that
this model gives statistically better fit compared to the other models, when all the
three GRBs are considered. Considering the fact that this model is also the preferred one
for the bright GRBs with single/ separable pulses (Basak \& Rao 2013), we can conclude 
that such a composite model needs to be examined for other GRBs too.
This spectral description has several attractive features
which can be used to constrain the GRB emission mechanism. 
The advantages of using the
 2BBPL model are:

\begin{enumerate}
\item{GRB~090902B is not unique: All the three bright LAT detected GRBs too are
consistent
with the  2BBPL model.}
\item{Physically reasonable non-thermal component: The data are consistent with
a power-law with the
same index for a given GRB. Hence, phenomenological explanation in terms of
non-thermal
phenomena like shock acceleration etc. are easy to implement. }
\item{Well behaved variation in the thermal component, including the initial
parts
of the GRB pulses.}
\item{Non-thermal component closely matches with the LAT emission.}
\end{enumerate}

If the  2BBPL model is the correct description of a GRB, it has the following
implications:

First, the existence of two  closely correlated blackbody temperatures
(with a similar ratio of temperature for different GRBs) provides 
a unique handle to pin down the radiation mechanism. If they are due to two
distinct locations in the photosphere, variation of temperature
gives the cooling mechanism. If these two temperatures are due to two
co-existing glory of photons as per the cannon ball model (Dado et al. 2007),
one can identify the higher temperature  with the typical photon field 
in the pre-supernova region (a few eV boosted to $\sim$100 keV by the 
cannon ball with a large bulk Lorentz factor, $\Gamma_0$) and the lower
temperature could be the photon field generated by some other process like
bremsstrahlung.
A correct identification of these two temperatures will provide
a good handle on $\Gamma_0$ and measuring such parameters for the
different pulses of a given GRB can provide other useful jet parameters like
the beaming angle. 
In the fire-ball scenario, on the other hand, thermal emission in the photosphere
will have an angular dependence and a range of temperatures is expected and
hence the multicolour black body description (the mBBPL model) would be a better
alternative. As can be seen from Table 2, the mBBPL and the 2BBL models give 
almost equally satisfactory results.

Second, the non-thermal component can be described as a power-law with a
constant or slowly varying index, extending all the way to GeV energies (though
we cannot completely  rule out spectral breaks/ change in index from MeV to GeV
regions).
The bulk of the initial prompt energy is in the thermal components and a
smoothly varying non-thermal component across MeV to GeV range should be quite
easy to handle (see for example Barniol Duran \& Kumar 2011). In the case of
GRB~090902B, the average spectral index of the power law in the MeV region
(1.76$\pm$0.17) for the 2BBPL model agrees very well with the index obtained in the GeV region
(1.76), whereas for GRB~090926A the average index (1.65$\pm$0.35) differs from that
derived in the GeV region (2.03). Hence we cannot completely rule out a
spectral break from the MeV to the GeV region.

We would like to point out that there are several observational features in the 
GRB prompt emission which indicate the possibility of a separate thermal
component rather than a continuous temperature distribution.
Guiriec et al. (2011) found that GRB~100724B requires an additional thermal
component at $\sim$38 keV. Considering the fact that the E$_{peak}$ for this
GRB was reported to be 350 keV (which is equivalent to a thermal spectrum of kT
$\sim$117 keV, since the thermal spectrum peaks at 3 kT), the derived ratio of
temperatures ($\sim$3) is quite similar to what we found in the present work.
Shirasaki et al. (2008) identified a low energy component which can be
modelled as a black body in GRB~041006 and the variation of the peak energy in
the multiple components in the time resolved spectra are found to be related to
each other.
Preece  et al. (1996) analysed the low energy data from the BATSE spectroscopic 
detectors and identified a low energy component in 15\%  of the bursts.
It is quite conceivable that the existence of two black bodies with correlated
behaviour is quite generic in all GRBs and they become evident in the data of
bright GRBs with very high peak energy (so that both blackbodies 
are within $\sim$10 -- 1000 keV region).

 Finally, the utility of the classical Band spectrum can be understood as a
collection of evolving BBPL/  2BBPL spectra and a 
smoothly evolving thermal spectra is a good
approximation to the Band model.
The fact that for sources like GRB~090926A the Band model gives a good fit to
the time resolved data comparable to the  2BBPL model could be 
an indication  that the Band model effectively captures an evolving
blackbody spectrum (evolving within the time bin) better than a  2BBPL model
with a constant temperature (within the bin). An empirical description
of an evolving  2BBPL model and developing a spectral model with such empirical
description and testing them with the data could be a way to understand the
time resolved spectra of GRBs.

\begin{acknowledgements} This research has made use of data obtained through
the
HEASARC Online Service, provided by the NASA/GSFC, in support of NASA High
Energy
Astrophysics Programs. We thank the anonymous referee for very useful comments.

\end{acknowledgements}

\clearpage




\end{document}